\documentclass[runningheads]{llncs}
\usepackage{xcolor}
\usepackage{url}
\usepackage{hyperref}
\usepackage{graphicx}
\begin{document}
\title{Domain Adaptation for Dense Retrieval through Self-Supervision by Pseudo-Relevance Labeling}
%
%
\author{Minghan Li \and Eric Gaussier}

%
%
\institute{Univ. Grenoble Alpes, CNRS, LIG\\  Grenoble, France\\ \email{minghan.li@univ-grenoble-alpes.fr, eric.gaussier@imag.fr } 
}

\maketitle              
\begin{abstract}
Although neural information retrieval has witnessed great improvements, recent works showed that the generalization ability of dense retrieval models on target domains with different distributions is limited, which contrasts with the results obtained with interaction-based models. 
To address this issue, researchers have resorted to adversarial learning and query generation approaches; both approaches nevertheless resulted in limited improvements. In this paper, we propose to use a self-supervision approach in which pseudo-relevance labels are automatically generated on the target domain. 
To do so, we first use the standard BM25 model on the target domain to obtain a first ranking of documents, and then use the interaction-based model T53B to re-rank top documents. We further combine this approach with knowledge distillation relying on an interaction-based teacher model trained on the source domain.
Our experiments reveal that pseudo-relevance labeling using T53B and the MiniLM teacher performs on average better than other approaches and helps improve the state-of-the-art query generation approach GPL when it is fine-tuned on the pseudo-relevance labeled data. 

\end{abstract}
\section{Introduction}

Information retrieval (IR) is playing a pivotal role in our daily life due to data explosion. Traditional IR approaches like BM25 \cite{RobertsonZ09} compute a similarity between a query and a document on the sole basis of the terms common to both. As such, they are unable to handle semantic matching between different surface forms. Neural information retrieval, with the advent of deep neural networks, has greatly improved IR systems through models which can capture the semantics of each term and compare them even if their surface form differs. A popular model in both Natural Language Processing (NLP) and IR is BERT \cite{devlin2019bert}, which is based on transformers \cite{vaswani2017attention} and is pre-trained on large scale unlabeled collections through self-supervision; BERT can be used on a variety of downstream tasks through fine-tuning.  

Neural IR models can be roughly classified into two categories \cite{guo2020deep}: interaction-based and representation-based (also called \textit{dense retrieval}) approaches. Interaction-based models have been shown to perform better in average than dense retrieval models; on the other hand, dense retrieval models are faster than interaction-based models, since the document representations can be generated and stored in advance, and preferred if one needs to deploy a model at large scale. This said, recent studies like BEIR \cite{thakur2021beir} showed that dense retrieval models trained on a source domain generalize less well than traditional models as BM25 and interaction-based models on out-of-distribution (OOD) data sets. Although training on target data sets with gold labels is a standard process, the annotation required may be both time consuming and expensive so that this approach may not be applicable to many real world usages. 
It is thus important to address the issue at OOD scenarios for dense retrieval, which can be done by exploiting existing large scale source labeled data like MS MARCO \cite{bajaj2018ms}.

One of the goals of domain adaptation \cite{wang2018DA,wang2022generalizing} is to make a model that has been trained on one domain, called the source domain, to perform well on another domain, called the target domain, without using human labels on the latter. Recently, various domain adaption techniques for dense retrieval have been proposed. Domain generalization based on data generation is one type of approaches \cite{wang2022generalizing} which has been followed in Ma et al.~\cite{ma2021zero} through a model called QGen which generates queries for the target domain using a query generator trained on the source domain. Along the same line, GPL \cite{wang-etal-2022-gpl} uses hard negatives and knowledge distillation and obtain state-of-the-art results on a number of BEIR data sets. However, the created queries are synthetic and may not resemble real target queries. 
Another popular and widely used approach is based on domain adversarial learning \cite{wang2022generalizing}. Very recently, Xin et al.~\cite{xin2022zero} proposed a model called MoDIR which adversially trains a dense retrieval encoder to learn domain-invariant representations for dense retrieval. However, such a learning objective may produce a poor embedding space and lead to unstable performance \cite{wang-etal-2022-gpl,karouzos2021udalm}. 

In this paper, we address \textbf{do}main generalization for \textbf{d}ense
\textbf{re}trieval through \textbf{s}elf-\textbf{s}upervision by pseudo-relevance labeling (in short, DoDress).
We first aim to build pseudo-relevance labels on the target domain using both unsupervised models like BM25 as well as interaction-based models solely trained on the source domain as T53B \cite{nogueira2020document} which act as re-rankers. The rationale for using interaction-based models in this setting lies in the fact that these models have been showed to behave relatively well on OOD data sets \cite{thakur2021beir}. 
Note that the heavy T53B model is only used for producing pseudo-relevance labels before training the dense retrieval model so that the overall approach is still efficient during the online search stage. This method eliminates the requirement for human annotations and enables the model to use genuine queries and documents of the target domain. In addition, we investigate the use of knowledge distillation on top of the previous steps with the goal to further improve the final dense retrieval model on the target domain.

Our contributions are twofold: first, we propose to use both standard unsupervised IR models as well as interaction-based models trained on the source domain to produce pseudo-labels on the target domain; second, we further propose and evaluate knowledge distillation in order to improve the final dense retrieval model. Experiments demonstrate the efficacy of our approach; they show in particular that it helps to improve the SOTA approach GPL when it is fine-tuned on the pseudo-labeled data.

\section{Related Work}

Wang et al.~\cite{wang2022generalizing} present a survey paper about domain generalization on unseen domains. Domain generalization or adaptation can be categorized into three groups: data manipulation, representation learning and learning strategy. There are two kinds of techniques in the first group: data augmentation \cite{prakash2019structured,tobin2017domain,shankar2018generalizing,volpi2018generalizing} which is commonly used in image data (for example, altering the location, textual of objects and adding random noise),
and data generation \cite{rahman2019multi,qiao2020learning,zhang2018mixup} which uses some models to generate new data to train a model. Representation learning group has domain-invariant representation
learning (e.g., domain adversarial learning) \cite{blanchard2021domain,ganin2016domain,motiian2017unified} and feature disentanglement methods \cite{li2017deeper,nam2021reducing,liu2021learning}. The third group has several categories, for example ensemble learning\cite{mancini2018best,d2018domain}, meta-learning \cite{li2018learning,du2020learning} and self-supervised learning based approaches (e.g., the task of solving jigsaw puzzles) \cite{carlucci2019domain,jeon2021feature}.

Similar strategies, such as domain generalization or transfer learning, are put forth by researchers for information retrieval. A strategy similar to the one adopted in this study is described in \cite{mokrii2021systematic} which carries out a systematic evaluation of transfer ability of BERT-based neural ranking models. The authors additionally use BM25 to generate pseudo-relevance labels. They do not, however, focus on dense retrieval models which are known to require complex training methods and a large amount of data in a distinct situation \cite{gao2021condenser}. Besides, only using BM25 to obtain pseudo-relevance labels might be a weak solution.
For interaction-based models \cite{lee2019latent} or learning sentence embeddings \cite{gao2021simcse,wang2021tsdae}, some publications suggest self-supervised techniques. These methods are frequently used for pre-training, however they do not explicitly focus on domain generalization \cite{wang-etal-2022-gpl}. Ma et al.~\cite{ma2021zero} proposes QGen, a generation approach to zero-shot learning for first-stage dense passage retrieval that makes use of synthetic question generation, allowing the construction of arbitrarily large yet noisy question-passage relevance pairs that are domain specific, in an effort to overcome the challenge that neural retrieval models need a large supervised training set to outperform conventional term-based approaches. The documents used to generate queries are viewed as positive and other in-batch instances are viewed as negative. Concurrently, Liang et al.~\cite{liang2020embedding} consider the two-tower dense passage retrieval architecture. Given that labeled data can be difficult to obtain and that neural retrieval models need a huge amount of data to be trained, they also suggest using synthetic queries produced by a large sequence-to-sequence (seq2seq) model for unsupervised domain adaptation. These two papers show the effectiveness of the query generation approach, which is also used in the GPL model \cite{wang-etal-2022-gpl}. GPL relies on a pre-trained T5 encoder-decoder \cite{raffel2020exploring} to generate queries from input passages. The input passages are seen as positive passages whereas similar passages retrieved using an existing dense retrieval model are constituted by the (hard) negative passages. The Margin-MSE loss \cite{hofstatter2020improving} is used as knowledge distillation to teach the dense retrieval model to learn from an interaction-based model. Experimental results show a state-of-the-art effectiveness on several BEIR \cite{thakur2021beir} collections.

Researchers have also explored alternative strategies for domain adaptation of dense retrieval models. \cite{xin2022zero} proposed a momentum adversarial domain invariant representation learning
(MoDIR) approach, which introduces a momentum method to train a domain classifier that distinguishes source and target domains. The dense retrieval encoder is then trained in an adversarial manner to learn domain-invariant representations. 
A momentum queue that records embeddings from several prior batches is used in order to strike a balance between accuracy and efficiency \cite{xin2022zero}. This approach is used on a trained ANCE model \cite{xiong2020approximate}. The results vary from one data set to the other, with sometimes important improvements and sometimes marginal gains or losses. Karouzos et al.~\cite{karouzos2021udalm} proposed UDALM for domain adaptation for sentiment classification through multi-task learning. It simultaneously learns the objective of the Masked Language Modeling (MLM) task on the target domain and the task from the source labeled data. However, this strategy was not designed for dense retrieval and, as mentioned in \cite{wang-etal-2022-gpl}, it does not work well for dense retrieval.

In this paper, we propose to do domain adaptation for dense retrieval through self-supervision by pseudo-relevance labeling. Instead of merely employing BM25 to produce pseudo-relevance labels, we also apply the cutting-edge, domain-generalizable T53B interaction-based model \cite{nogueira2020document}. This model can produce more accurate pseudo-relevance labels, where top ranked documents are viewed as relevant to a given query. Additionally, knowledge distillation is employed to improve the model effectiveness while not sacrificing to its efficiency during retrieval.

\section{Background}

\paragraph{Dense retrieval} \cite{karpukhin2020dense,xin2022zero} seeks to encode both queries and documents into a low-dimensional space with an encoder $g$, typically a BERT-like model. The retrieval status value (RSV) of a query and a document is then calculated with a simple similarity function in the low-dimensional space:
\begin{equation}
    RSV(q,d)_{DR} = g(q) \cdot g(d)\  \,\, \left( or \,\, RSV(q,d)_{DR} = cos(g(q),g(d)) \right), \nonumber
\end{equation}
where $g(q)$ (resp. $g(d)$) denotes the encoding of the query (resp. document).
This enables a fast retrieval through a nearest neighbour search strategy \cite{xiong2020approximate}.

\paragraph{BM25} BM25 is a widely used standard IR algorithm based on term matching. The RSV of a document with respect to a query is given by:
\begin{equation}
    RSV(q,d)_{BM25} = \sum_{w \in q \cap d} IDF(w) \cdot \frac{tf_{w}}{k_1 \cdot (1-b+b\cdot \frac{l_{d}}{l_{avg}})+tf_{w}}, \nonumber
\end{equation}
where $IDF(w)$ is the inverse document frequency, $l_{d}$ is the length of document $d$, $l_{avg}$ the average length of the documents in the data set, and $k_1$ and $b$ two hyper-parameters

\paragraph{T53B} By establishing a uniform framework that transforms all text-based language problems into a text-to-text format, T5 \cite{raffel2020exploring} explores the landscape of transfer learning for NLP and achieves state-of-the-art results on many benchmarks. Nogueira et al.~\cite{nogueira2020document} proposed to use T5 as an interaction-based model for information retrieval by relying on the following input representation:
\begin{center}
\textit{Query: [q] Document: [d] Relevant: true or false}
\end{center}
\noindent where $[q]$ and $[d]$ are replaced with the query and document texts. During training, the T5 model learns to generate the word ``true'' when the document is relevant to the query, and the word ``false'' when it is not. The relevance score for inference is then determined by the likelihood of producing ``true'' \cite{nogueira2020document}:
\begin{equation}
    RSV(q,d)_{T5} = softmax(Z_{true}) = \frac{e^{Z_{true}}}{e^{Z_{true}}+e^{Z_{false}}}, \nonumber
\end{equation}
where $Z_{true}$ and $Z_{false}$ are the logits of output tokens.

\section{DoDress: Pseudo-Relevance Label Generation and Knowledge Distillation}
\label{secMethod}

\begin{figure}
\centering
\includegraphics[width=250pt]{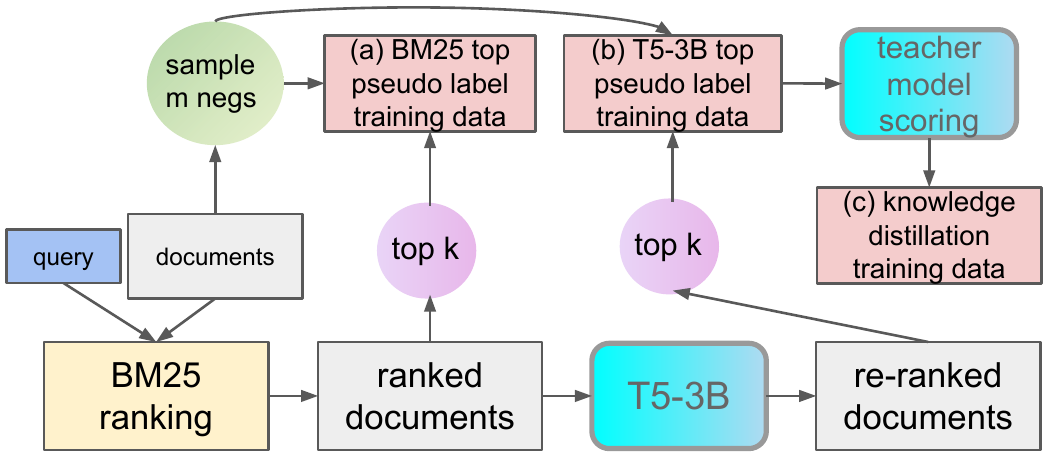}
\caption{The overall pipeline of generating different self-supervised data with pseudo-relevance labeling.} \label{figArch}
\end{figure}

Our approach is described in Figure~\ref{figArch}, where (a), (b), and (c) represent the three different training data generated on the target corpus.

\subsection{Generating Positive-Negative Training Pairs on the Target Domain}
\label{pairgeneSec}

We simply propose here to consider the top $k$ documents, obtained with BM25 or with the combination BM25\&T53B in which T53B serves as a re-ranker, as relevant. $k$ is an hyper-parameter which can be set according to different information, as, \textit{e.g.}, the number of available queries and documents. Furthermore, for each pseudo-relevant query-document pair, we randomly sample $m$ documents from the target collection and consider them as non-relevant. Thus, for each query, $k\times m$ query-document triplets (query, relevant document, non-relevant document) can be formed. The training data (a) and (b) in Figure~\ref{figArch} correspond to the data generated by the above process with BM25 for (a) and BM25\&T53B for (b).

\subsubsection{Pairwise Loss} In this paper, we rely on the RankNet pairwise loss \cite{burges2010ranknet,li2022bert} to train a dense retrieval model using the triplets generated above, defined by:
\begin{equation}
  \mathcal{L}(q,d^+,d^-;\Theta) = -\log(\sigma(S_{q,d^+}-S_{q,d^-})), \nonumber
\end{equation}
where $q$ is a query, $(d^+_q,d^-_q)$ is a (positive,negative) training document pair for $q$, $\sigma$ is the sigmoid function, $\Theta$ represents the parameters of the dense retrieval model, and $S_{q,d}$ is the score provided by the model for document $d$ with respect to query $q$. 

\subsection{Learning the Dense Retrieval Model through Knowledge Distillation}

In knowledge distillation, a (usually complex) teacher model teaches a (usually simpler) student model to mimic its behaviour through an appropriate loss function. As mentioned before, interaction-based models perform better than dense retrieval models in OOD settings. Thus, rather than simply learning the dense retrieval model for the target domain using the triplets generated above and the Ranknet pairwise loss, we propose to learn the dense retrieval model on the above triplets using knowledge distillation, where an interaction-based model is used as the teacher model and where the dense retrieval plays the role of the student model. This can be done, following \cite{wang-etal-2022-gpl,hofstatter2020improving}, by considering the MarginMSE loss, defined by:
\begin{equation}
  \mathcal{L}(q,d^+,d^-;\Theta) = MSE(S_{q,d^+}^s-S_{q,d^-}^s, S_{q,d^+}^t-S_{q,d^-}^t)
  , \nonumber
\end{equation}
where $S_{q,d}^s$ is the score provided by the student model and $S_{q,d}^t$ the score provided by the teacher model. As before, $\Theta$ represents the parameters of the student dense retrieval model. As can note, this loss encourages the student model to imitate the teacher model's output on the triplets generated above. Note that the teacher is only trained on the source data.

A natural choice for the teacher model is T53B which is considered as the interaction-based model that generalizes best on OOD data. During preliminary experiments, however, we found that this teacher model is sometimes unstable, mainly due to the large differences between the teacher and student models. We thus make use in our experiments of an additional teacher, namely \textit{ms-marco-MiniLM-L-6-v2} which is also used in GPL \cite{wang-etal-2022-gpl}.

\subsection{Improving GPL: Combining Pseudo-Relevance Labels and Pseudo-Queries}

As mentioned before, the QGen and GPL approaches both rely on a query generator to generate pseudo-queries in order to train a dense retrieval model. We propose here to further train this dense retrieval model on the pseudo-relevance triplets described in Section~\ref{pairgeneSec}. We believe one can gain from this additional training on the target collection as pseudo-queries and pseudo-relevance labels rely on different sources of information and are complementary to each other. As we will see in the experimental section, this combination indeed improves the pseudo-query generation approach.

\section{Experiments}

\subsection{Data Sets}

The MS MARCO passage ranking data set ~\cite{bajaj2018ms} is used as the source domain data. We experiment on 6 target domain data sets from the BEIR benchmark \cite{thakur2021beir}. They are FiQA, finance question answering \cite{fiqa-2018}, SciFact, scientific fact checking \cite{scifact-2020}, BioASQ biomedical question answering \cite{bioasq-2015} (following \cite{wang-etal-2022-gpl}, irrelevant documents are randomly eliminated, leaving 1M documents), TREC-COVID, bio-medical papers \cite{trec-covid-2020} (following \cite{wang-etal-2022-gpl}, documents with only title and no body contents are deleted), Touché-2020, argument retrieval, miscellaneous domain \cite{10.1007/978-3-030-58219-7_26}, and Robust04, news documents \cite{robust04-2005}. Different topics and tasks are covered by these chosen data sets.

\subsection{Experimental Setting}

Our implementation is based on the matchmaker open-source framework\footnote{\url{https://github.com/sebastian-hofstaetter/matchmaker}} which is modified with mean pooling and dense evaluation\footnote{\url{https://github.com/UKPLab/gpl/blob/main/gpl/toolkit/evaluation.py}} using automatic mixed precision \cite{micikevicius2018mixed}.
On the target collection, the training triplets are generated according to the approaches described in Section \ref{secMethod}, either using BM25 or the combination BM25+T5 to produce pseudo-relevant documents by considering the top $k$ documents as relevant. Following previous work and for fair comparison, the transformer architecture of the dense retrieval model, referred to as D-BERT, is DistilBERT~\cite{DBLP:journals/corr/abs-1910-01108} with 6 layers. For domain adaptation, D-BERT is firstly trained on the source domain. We conduct two groups of experiments. For the first group, the dense retrieval models D-BERT and GPL are trained using the RankNet pairwise loss on both sets of triplets (obtained by either BM25 or BM25+T5).  Note that GPL is first trained on the target queries and associated relevant documents it generates prior to be trained on the target triplets. For the second group, we both train D-BERT and GPL on the triplets obtained with BM25+T5 using knowledge distillation and the MarginMSE loss considering D-BERT and GPL as students and either T5 or MiniLM as teacher. As before, GPL is first trained on the target queries and associated relevant documents it generates prior to be trained on the target triplets. The T5 model used is the 3B version that is trained on MS MARCO passage ranking data set\footnote{\url{https://huggingface.co/castorini/monoT53B-D-BERT-10k}}. The MiniLM cross-encoder used is the \textit{ms-marco-MiniLM-L-6-v2} version\footnote{\url{https://huggingface.co/cross-encoder/ms-marco-MiniLM-L-6-v2}} from Sentence Transformers \cite{reimers2019sentence}.

For constructing the training set, we select the number $k$ of top documents to be considered as relevant according to the number of queries (to generate enough pairs) and documents (as a large number of documents enables sampling more negative documents). For each relevant document, we randomly select $m$ documents from the target collection not present in the top $k$ list of the query. These documents are considered as not relevant. Table \ref{tab.stat} displays the number of queries, the value selected for $k$ (in parenthesis) and the number $m$ of non relevant  documents per relevant document. For BioASQ for example, the number of triplets in the training set is $450\times 2 \times 100 = 90000$. At the end, each data set has a number of triplets in the training set in the range of 50000 to 100000. We also construct a development set to select the hyper-parameters of the models on each collection. For each query, the top 10 documents are considered relevant and 90 randomly selected documents (not in the top 10) as non relevant. This choice is dictated by the fact that we need a sufficient number of relevant documents for evaluation purposes and that we have a limited number of queries for the development set. However, to counterbalance the risk of considering as relevant documents which in fact are not, the top two documents are labeled as '2' and the following eight ones as '1'. The non-relevant documents are labeled as '0', thus leading to 3-level relevance judgements for each data set. The best model is saved according to the NDCG@10 score on the development set evaluated every 1K steps.

Following \cite{wang-etal-2022-gpl}, a maximum sequence length of 350 with mean pooling and dot-product similarity is used. For all data sets, we use a batch size of 8, which means 8 positive-negative pairs, and a learning rate of 2e-6 with Adam optimizer for 10K training steps. Cosine LR schedule \cite{loshchilov2017sgdr} is also used for learning rate decay.

\begin{table}
\caption{The top $k$ selected as positive and $m$ as negative for each data set. The number in parentheses is used for generating training data, remaining for Dev set.}\label{tab.stat}
\centering
\scalebox{0.9}{
\begin{tabular}{|l|l|l|l|l|l|}
\hline
data set &  \#queries & \#docs  & top $k$ as relevant & $m$ as non-relevant\\
\hline
FiQA & 6648 (6598) & 57K    &1 & 10\\
SciFact &  1109 (1059) &5K& 5 &10\\
BioASQ & 500 (450) &1M& 2 & 100\\
TRECC. & 50 (40) &129K& 10 &150\\
Touché-2020 & 49 (39) &382K&  10 & 150\\
Robust04 & 250 (200) &528K&  10 & 50\\
\hline
\end{tabular}
}
\end{table}

\subsection{Baselines}

Following \cite{wang-etal-2022-gpl}, we compare the proposed approaches with zero-shot models, with pre-training approaches and with recent SOTA approaches to domain adaptation. 

\subsubsection{Zero-shot models} Baseline zero-shot models comprise BM25 based on Anserini \cite{yang2018anserini} with default parameters which obtains top 100 documents for each query and does not require to be trained is compared (these BM25 ranking lists are further used to generate pseudo-labeling training data in this paper) and the dense retrieval model D-BERT 
solely trained on the source collection with hard negatives and the MarginMSE loss with the \textit{ms-marco-MiniLM-L-6-v2} model taken as the teacher model (it is further used as a start point for domain adaptation). 

\subsubsection{Pre-trained models} We compare with SimCSE \cite{gao2021simcse}, ICT \cite{lee2019latent} and TSDAE \cite{wang2021tsdae}. These models are all firstly pre-trained in a self-supervised way on the target data set and then fine-tuned on MS MARCO.

\subsubsection{Domain adaptation approaches}
We compare here with four recent SOTA approaches: MoDIR \cite{xin2022zero} which is based on ANCE \cite{xiong2020approximate} and relies on an adversarial training, UDALM \cite{karouzos2021udalm} which is based on multi-task learning, and QGen \cite{ma2021zero} and GPL \cite{wang-etal-2022-gpl} which are approaches based on query generation. 

\subsubsection{} In addition, we make use of the interaction-based models BM25+CE and BM25+T53B which can be seen as strong baselines due to the good behaviour of interaction-based models in OOD settings \cite{thakur2021beir} but which are nevertheless inefficient at inference. These models re-rank the top 100 BM25 ranked list, using a \textit{ms-marco-MiniLM-L-6-v2} and T53B cross encoders respectively. 

\subsection{Results and Analysis}

Table \ref{tbl:main_results} displays the results obtained with the different models and approaches. The results reported for BM25+CE, UDALM, MoDIR, SimCSE, ICT, TDSAE, QGen and TSDAE+GPL are from \cite{wang-etal-2022-gpl}. The notation ``DoDress-BM25 (D-BERT)'' (respectively ``DoDress-T53B (D-BERT)'') corresponds to the D-BERT dense retrieval model pre-trained on MS MARCO and fine-tuned on the target data using the pseudo-relevance labels generated using BM25 (respectively using BM25+T5). The notation (GPL) means the same for GPL, which is again first trained on the target queries and associated relevant documents it generates prior to be trained on the target triplets. The notation ``DoDress-teach-D (stu)'' where `teach' is either T53B or MiniLM and `stu' is either D-BERT or GPL corresponds to the scenraio with knowledge distillation.

We analyze the results by answering three research questions.

\begin{itemize}
    \item[\textbf{RQ1}] Do BM25 top positives help domain generalization for dense retrieval?
\end{itemize}

As one can see in Table~\ref{tbl:main_results}, DoDress-BM25 (D-BERT) improves over D-BERT on four data sets out of six, the two data sets on which it does not perform well being FiQA and TREC-COVID. DoDress-BM25 (GPL) shows a similar trend compared to GPL. On Touché-2020, DoDress-BM25 (D-BERT) shows a 43.4\% ($(28.1-19.6)\div 19.6$) improvement compared with D-BERT. These results show that although the pseudo-relevance labels obtained with BM25 can help generalize to new domains, the approach is not entirely stable, which is certainly due to incorrect pseudo-relevance labels.

\begin{itemize}
    \item[\textbf{RQ2}] Does the re-ranking of top positives with T53B help achieve better domain generalization for dense retrieval?
\end{itemize}

Compared with D-BERT, DoDress-T53B (D-BERT) shows better NDCG@10 results on all six data sets. DoDress-T53B (GPL) has a similar trend and yields positive domain generalization effects on all six data sets. For example, on Robust04, DoDress-T53B (GPL) achieved 45.1 on NDCG@10, further improving the SOTA approach GPL with a 8.7\% improvement. On average, DoDress-T53B (GPL) achieves the best results, and outperforms TSDAE + GPL. DoDress-T53B (GPL) is furthermore the only dense retrieval model that outperforms in average BM25, this latter model having a very good average performance due to its exceptional result on Touché-2020,

All in all, the strategy of re-ranking BM25 top documents with T53B is beneficial on all collections for both dense models D-BERT and GPL. It is more accurate than the strategy based on BM25 only on all collections but on Touché-2020. The best model on this latter collection is by far BM25 (Anserini), used in a zero-shot manner. Using only the pseudo-relevance labels from BM25 on this collection yields an annotation in line with this best model, which explains why the strategy BM25+T53B does not perform as well as the simpler BM25 strategy on this collection. 

\begin{itemize}
    \item[\textbf{RQ3}] Can knowledge distillation further improve  domain generalization?
\end{itemize}

As one can note, DoDress-T53B-D achieves better results than DoDress-T53B on several data sets. For example, on Robust04, DoDress-T53B-D (D-BERT) achieves 45.2 on NDCG@10, a 19.6\% improvement over D-BERT (and a score even better than DoDress-T53B (GPL)). However, on several data sets, this approach fails and even leads to worse results as on TREC-COVID. We hypothesize this is due to the difference between the teacher and student models, as T53B contains 3 billion parameters while the dense retrieval model only contains 66 million parameters. 

Using MiniLM as teacher which contains 22 million parameters, we can see that DoDress-MiniLM-D (D-BERT) obtains better results than DoDress-T53B (D-BERT) on four data sets out of six, and better results than DoDress-T53B-D (D-BERT) on five data sets out of six. DoDress-MiniLM-D (GPL) shows a similar trend, except on Robust04. On average, DoDress-MiniLM-D (D-BERT) achieves the best results for the D-BERT based models while DoDress-MiniLM-D (GPL) is the scond best model for GPL, just below DoDress-T53B (GPL) which benefits from a higher score on Robust04. These results suggest that knowledge distillation based on a teacher model as the cross-encoder model MiniLM helps domain generalization of dense retrieval models. 

In conclusion, the above results demonstrate the effectiveness of the proposed approach which provides state-of-the-art generalization results of dense retrieval models on several collections.

\begin{table}
\caption{Evaluation using nDCG@10 (\%). TRECC and Touché are short for TREC-COVID and Touché-2020. Results of positive domain generalization effects of proposed approach DoDress compared with D-BERT and GPL respectively are \textbf{bold}.
}
\centering
\scalebox{0.88}{
\begin{tabular}{|l|c|c|c|c|c|c|c|} 
\hline
\textbf{Method}
& \textbf{FiQA}          & \textbf{SciFact}       & \textbf{BioASQ} & \textbf{TRECC}        & \textbf{Touché}       & \textbf{Robust04}      & \textbf{Avg.}        \\ 
\hline
\multicolumn{8}{|l|}{\textit{Zero-Shot Models}}                                                                                                                                                                      \\ 
\hline
D-BERT                & 26.7 & 57.1 & 53.6 & 65.1 & 19.6 & 37.8 & 43.4 \\
BM25 (Anserini)     &23.6 &67.9 &73.0 &58.5 & 44.2 &44.9 & 52.0     \\
\hline

\multicolumn{8}{|l|}{\textit{Re-Ranking with Cross-Encoders (Upper Bound, Inefficient at Inference)}}                                                                                                                                                        \\ 
\hline
BM25 + CE   & 33.1 & 67.6 & 72.8 & 71.2 & 27.1 & 46.7 & 53.1 \\
BM25 + T53B   & 39.2 & 73.1 & 76.1 & 78.0 & 30.6 & 50.4 & 57.9 \\
\hline
\multicolumn{8}{|l|}{\textit{Previous Domain Adaptation Methods}}                                                                                                                                                                       \\ 
\hline

UDALM                                      & 23.3 & 33.6 & 33.1 & 57.1 & - & 26.3 & -- \\
MoDIR (ANCE)                                     & 29.6 & 50.2 & 47.9 & 66.0 & 31.5 & -- & -- \\
\hline
\multicolumn{8}{|l|}{\textit{Pre-Training based Domain Adaptation: Target $\to$ D-BERT}}                                                                                                                           \\ 
\hline
SimCSE & 26.7 & 55.0 & 53.2 & 68.3 & - & 37.9 & - \\
ICT & 27.0 & 58.3 & 55.3 & 69.7 & - & 37.4 & - \\
TSDAE & 29.3 & 62.8 & 55.5 & 76.1 & 21.8 & 39.4 & 47.5 \\
\hline
\multicolumn{8}{|l|}{\textit{Generation-based Domain Adaptation (Previous State-of-the-Art)}}                                                                                                                        \\ 
\hline
QGen    & 28.7 & 63.8 & 56.5 & 72.4 & 17.1 & 38.1 & 46.1 \\
GPL & 32.8 & 66.4 & 62.8 & 72.6 & 23.1 & 41.5 & 49.9 \\
TSDAE + GPL &  34.4 & 68.9 & 61.6 & 74.6 & 23.5 & 43.0 & 51.0 \\
\hline

\multicolumn{8}{|l|}{Proposed Method: BM25 Top as Positive}                                                                                                                                          \\ 
\hline
DoDress-BM25 (D-BERT) & 23.3 & \textbf{59.3} & \textbf{53.9} & 64.5 & \textbf{28.1} & \textbf{39.4} & \textbf{44.75} \\
DoDress-BM25 (GPL) &  31.3 & \textbf{68.6} & \textbf{63.1} & 69.0 & \textbf{29.9} & \textbf{43.4} & \textbf{50.9} \\
\hline

\multicolumn{8}{|l|}{Proposed Method: BM25, T53B re-rank Top as Positive}                                                                                                                                           \\ 
\hline
DoDress-T53B (D-BERT) & \textbf{28.0} & \textbf{62.3} & \textbf{56.4} & \textbf{70.7} & \textbf{23.3} & \textbf{41.0} & \textbf{47.0} \\
DoDress-T53B (GPL) &  \textbf{33.0} & \textbf{68.8} & \textbf{64.1} & \textbf{76.5} & \textbf{25.9}& \textbf{45.1} & \textbf{52.2}\\
\hline

\multicolumn{8}{|l|}{Proposed Method: BM25, T53B re-rank Top as Positive, Distillation}                                                                                                                                           \\ 
\hline
DoDress-T53B-D (D-BERT) & 26.2 & \textbf{63.4} & \textbf{57.1} & 36.9 & \textbf{21.1} & \textbf{45.2} & 41.7 \\
DoDress-T53B-D (GPL) &  \textbf{33.2} & \textbf{67.3} & \textbf{64.1} & 45.7 & \textbf{24.9} & \textbf{44.9} & 46.7 \\
\hline
DoDress-MiniLM-D (D-BERT) & \textbf{29.5}   & \textbf{64.5} & \textbf{61.7} & \textbf{69.7}  & \textbf{24.0} & \textbf{39.7} & \textbf{48.2} \\
DoDress-MiniLM-D (GPL) & \textbf{33.0}  &  \textbf{68.2} & \textbf{65.7} &  \textbf{76.6} & \textbf{26.2} & 41.1 & \textbf{51.8} \\
\hline
\end{tabular}
}
\vspace{-0.3cm}
\label{tbl:main_results}
\end{table}

\subsection{Different Top $k$ Positives}

\begin{figure}
\centering
\includegraphics[width=180pt]{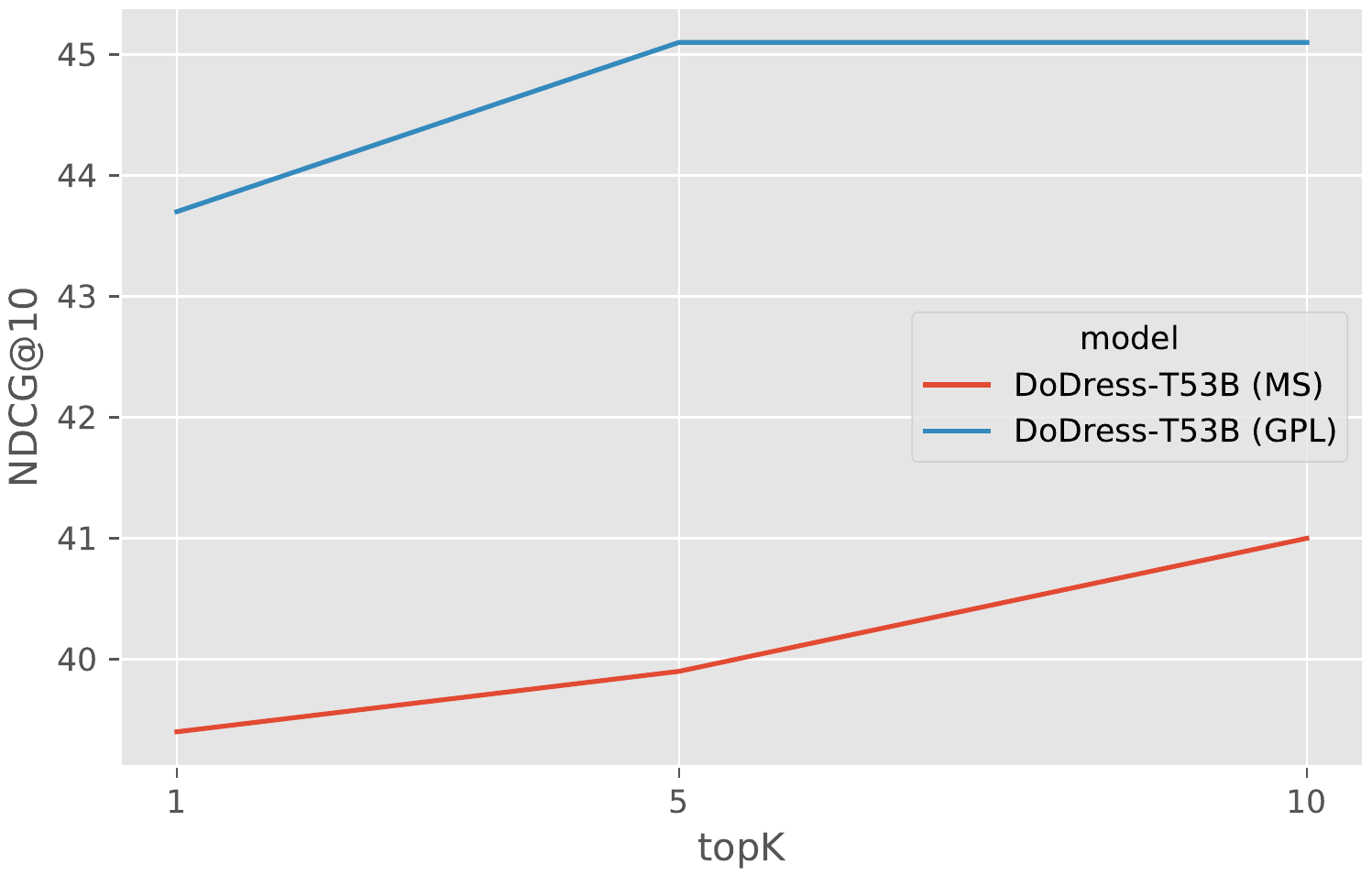}
\caption{NDCG@10 (\%) scores of different top $k$ selected as positive pseudo-labels on Robust04 data set.} \label{fig1}
\vspace{-1cm}
\end{figure}

We also conducted an experiment aiming to assess the impact of the number $k$ of top documents retained as relevant on Robust04. Figure~\ref{fig1} shows the NDCG@10 scores for $k \in \{1,5,10\}$ using the two models DoDress-T53B (D-BERT) and DoDress-T53B (GPL). Each relevant document is matched with respectively 500, 100, 50 randomly selected non relevant documents ($m$ values). As one can see, the best results are obtained with $k=10$, even though there is no difference between 5 and 10 for DoDress-T53B (GPL). We conducted the same experiment on BioASQ, with $k \in \{2,5\}$ and respectively using 100 and 40 non relevant documents. In this case, the NDCG@10 score for DoDress-T53B (D-BERT) is 56.1 for $k=2$, while the score is 55.0 for $k=5$. 

These results show that the choice of $k$ is important and that it should depend on the collection considered. Our strategy in this study has been to  select the value of $k$ according to the number of available queries by using larger values of $k$ when the number of available queries number is smaller, which can be seen as a conservative strategy for collections with a large number of queries. This said, other strategies need be explored in the future.

\section{Conclusion}

This paper studies whether one can benefit from existing IR models, either unsupervised or pre-trained on MS MARCO, to generate pseudo-relevance labels on an unannotated target collection. These labels are then used to fine-tune dense retrieval models on the target collection. Our study reveals that this approach works well when the pseudo-labels are generated using a combination of BM25 and T53B, and that it helps improve the generalization results of the GPL model which also makes use of generated queries and associated relevant documents on the target collection.

We further study the combination of this approach with knowledge distillation, in which an interaction-based model is use as the teacher model, the student model being the dense retrieval model. Our experiments indicate that the proposed pseudo-labeling with T53B and using MiniLM teacher outperform other methods on average, obtaining new state-of-the-art results on several data sets for DR model's domain generalization.

%
%
%
\bibliographystyle{splncs04}
\bibliography{mybibliography}
%




\end{document}